\pdfoutput=1

\documentclass[gmd, manuscript]{copernicus}

\begin{document}
\nolinenumbers   

\title{Hydrograd v2.0: a differentiable two-dimensional shallow-water model on unstructured meshes with steady and unsteady adjoints for calibration and closure learning}
 
\Author[1,2][xzl123@psu.edu]{Xiaofeng}{Liu}

\affil[1]{Department of Civil and Environmental Engineering, The Pennsylvania State University, University Park, PA 16802, USA}
\affil[2]{Institute for Computational and Data Sciences, The Pennsylvania State University, University Park, PA 16802, USA}

\runningtitle{Hydrograd v2.0: differentiable shallow-water modelling on unstructured meshes}

\runningauthor{X. Liu}

\received{}
\pubdiscuss{}
\revised{}
\accepted{}
\published{}

\firstpage{1}

\maketitle

\begin{abstract}
We present \textit{Hydrograd} v2.0, a significantly improved differentiable 2D shallow-water model written in Python with JAX. The model solves the shallow-water equations on the unstructured meshes of production models such as SRH-2D and HEC-RAS 2D with a well-balanced, positivity-preserving finite-volume scheme that handles wetting and drying. Gradients are tailored to the type of problem. Steady states are found by Newton's method with an exact sparse Jacobian obtained by graph colouring, and their gradients follow from the implicit function theorem at the cost of one linear solve. Unsteady flows are differentiated in reverse mode with checkpointing, or with an implicit time integrator whose gradient costs about two forward runs. All gradients agree with finite differences, to about $10^{-8}$ for steady flows. On the Savannah River case, the steady water surface differs from that of SRH-2D by 4.7\,mm (root mean square). Steady closure learning was 96 times faster than differentiating through time marching and reached a lower loss. An inflow hydrograph was inverted from stage records with a peak error of 2\,\%. In an unsteady roughness calibration, a singular value decomposition analysis of the sensitivity matrix showed that the stage records at the gauges are 70 to 160 times more sensitive to the total resistance of the reach than to its distribution among the roughness zones, which explains why only a gauge layout with gauges in every zone recovered all five roughness values. On one GPU, a mesh of 512,000 cells takes 7.6\,s for a forward run and 51\,s for a gradient.
\end{abstract}  

\introduction

Rivers and their floodplains convey water, sediment, nutrients and carbon from the land to the oceans, and floods are among the most damaging natural hazards. Two-dimensional (2D) depth-averaged models based on the shallow-water equations (SWEs) are the standard tools for simulating these flows. They are used for flood-inundation mapping \citep{Bates2010,Shaw2021}, river morphodynamics and ecohydraulics \citep{Radiyan2026}, and in engineering practice. Models such as SRH-2D \citep{Lai2010}, HEC-RAS 2D \citep{Brunner2016} and TELEMAC-2D \citep{Hervouet2007} solve the SWEs on unstructured meshes, which follow river banks, levees and structures. One advantage of unstructured mesh is that the models can concentrate resolution where the flow changes rapidly. In all of these models, many sub-grid processes, such as bedforms and vegetation induced flow resistance, have to be parameterised, much like turbulence and convection in atmospheric and ocean models. Flow resistance is usually represented by Manning's coefficient $n$, assigned zone by zone or through empirical laws of the flow depth, velocity and roughness height \citep{Liu2024NCHRP}. Flow resistance parameters cannot be measured directly, so it is inferred by fitting simulated water-surface elevations (WSEs), velocities or inundation extents to observations. Calibration can also be done on other inputs to the model, such as the inflow hydrograph or the bathymetry, which are often uncertain.

Many calibrations are done manually, with model-independent tools such as PEST, which obtain derivatives by finite differences \citep{Doherty2015}, or with Monte Carlo methods such as GLUE \citep{Pappenberger2005}. These approaches have shown that model parameters are often poorly identified, with many parameter sets fitting the data equally well. Their cost also grows quickly with the number of parameters, so calibrated models often can only have a few parameters. As a results, spatially distributed or flow-dependent resistance is rarely inferred from data. Meanwhile, the observations of rivers are growing rapidly, from dense gauge networks and image-based velocimetry \citep{Tenorio2026} to satellite altimetry. The Surface Water and Ocean Topography (SWOT) mission now measures the elevation, width and slope of rivers worldwide \citep{Biancamaria2016}, but discharge estimated from such data relies on prior information and is often biased \citep{Durand2016}. 

Recently, data-driven closures have gained significant popularity in computational modeling of complex dynamical systems. Instead of prescribing an empirical law for a poorly understood sub-grid process, the process is represented by a neural network (NN) that is embedded in the governing equations and learned from data. For computational hydraulics, \citet{Liu2025} proposed the universal shallow-water equations (USWEs), in which the flow resistance closure of the SWEs is replaced by an NN. The NN takes local flow variables such as the depth and velocity as inputs and returns the resistance coefficient, so the resistance can vary in space and with the flow condition without a prescribed functional form. Mass and momentum conservation remain enforced, and only the closure is learned. The difficulty lies again in training. An NN closure has hundreds to thousands of weights, far more than the handful of zonal Manning's coefficients that the calibration methods above can handle. Derivative-free and sampling methods are impractical in a parameter space of this size. The weights can only be found with the gradients of the model output with respect to every weight, which requires the hydrodynamic model itself to be differentiable \citep{Shen2023}.

Gradient-based inference has a long history in the geosciences in the form of adjoint models for data assimilation \citep{Errico1997}. For the SWEs, adjoint codes were obtained by source transformation of Fortran solvers, e.g. DassFlow \citep{Monnier2016} and TELEMAC with algorithmic differentiation \citep{Villaret2016}, or from high-level finite-element frameworks, e.g. Thetis with dolfin-adjoint \citep{Karna2023}. More recently, automatic differentiation (AD) in machine-learning frameworks has made it possible to differentiate a whole model and to embed NNs in it, an approach known as differentiable modelling \citep{Shen2023,Gelbrecht2023,Sapienza2024}. Hybrid models of this kind have learned parameterisations of hydrological processes from large data sets \citep{Tsai2021,Song2024} and of the atmosphere \citep{Kochkov2024}. For 2D river hydrodynamics, the USWEs of \citet{Liu2025} were trained with a differentiable solver, \textit{Hydrograd} v1.0 (Hydrograd.jl), written in Julia within the scientific machine learning (SciML) ecosystem \citep{Rackauckas2020}.

\textit{Hydrograd} v1.0 demonstrated the concept, but with several limitations. First, steady flows, the most common calibration target, were obtained by marching the unsteady equations in time and differentiating through every time step. The gradient then cost many forward runs. When the march had not reached steady state, the gradient depended on the final simulated state rather than on the true steady state itself. Second, the Roe-type scheme \citep{Roe1981} was not designed for wetting and drying, which excluded floodplain inundation. Third, the memory needed by reverse-mode AD grows with the number of time steps, which limits the length of unsteady simulations. Fourth, the code ran mostly on central processing units (CPUs).

Other differentiable SWE solvers have appeared recently \citep{Li2026Inunda,Xu2026DiffSWE,CardosoBihlo2026}. Like the general-purpose differentiable flow solvers JAX-Fluids \citep{Bezgin2023} and JAX-CFD \citep{Kochkov2021}, they are written in machine-learning frameworks that run on graphics processing units (GPUs), but they use structured Cartesian grids. A structured grid suits the data layout of a GPU, but it cannot follow the geometry of real rivers without very fine resolution everywhere, and it does not match the meshes that modellers already build for SRH-2D, HEC-RAS 2D or TELEMAC. Differentiable finite-volume solvers for unstructured meshes have only just appeared, for incompressible and compressible flows \citep{Du2027DiFVM,deRomemont2026}, not for the SWEs. Conversely, the SWE adjoint models that do work on unstructured meshes run on CPUs and were not designed to train data-driven closures \citep{Monnier2016,Villaret2016,Karna2023}. In the author's opinion, a differentiable SWE model that is to be used for real rivers and floodplains must combine the unstructured meshes of production models with the efficiency of modern AD on GPUs, and it must provide gradients whose cost matches the type of problem being solved.

In this work, we present \textit{Hydrograd} v2.0, a re-designed and extended version of \textit{Hydrograd} v1.0 \citep{Liu2025}, written in Python with JAX \citep{Bradbury2018}. The contributions of this work are as follows. (i) The model solves the SWEs with a well-balanced, positivity-preserving finite-volume scheme on the unstructured meshes, including wetting and drying, at first or second order. Every part of the model, including the exact Jacobian matrix of the discrete equations, runs on these meshes. (ii) For steady flows, the model finds the steady state with Newton's method and computes gradients with the implicit function theorem: one additional sparse linear solve, independent of the number of parameters and of the Newton's iterations to the steady state. (iii) For unsteady flows, the model offers reverse-mode AD with checkpointing for explicit time stepping, and an implicit time integrator whose gradient requires one linear solve per time step, so that long simulations can be differentiated at a cost comparable to the forward run. (iv) The model includes differentiable closures for flow resistance (Manning laws and NN closures) and for horizontal turbulent mixing, and it runs on CPUs and GPUs. We verify the model against analytical solutions, evaluate it against SRH-2D, and demonstrate it on roughness calibration, closure learning and hydrograph inversion.

The rest of the paper is organised as follows. Section~\ref{sec:model} describes the governing equations, the closures, the discretisation on unstructured meshes, the solvers and the computation of gradients. Section~\ref{sec:verification} verifies the numerical solutions against analytical solutions. Section~\ref{sec:applications} evaluates the model and demonstrates its new capabilities in steady and unsteady applications, verifies the gradients of every method against finite differences, and reports its computational performance. Section~\ref{sec:discussion} discusses the choice of gradient method, the scope and the limitations of the model, and Section~\ref{sec:conclusions} concludes the paper.

\section{Model description}
\label{sec:model}

\begin{figure*}[htp]
\includegraphics[width=17cm]{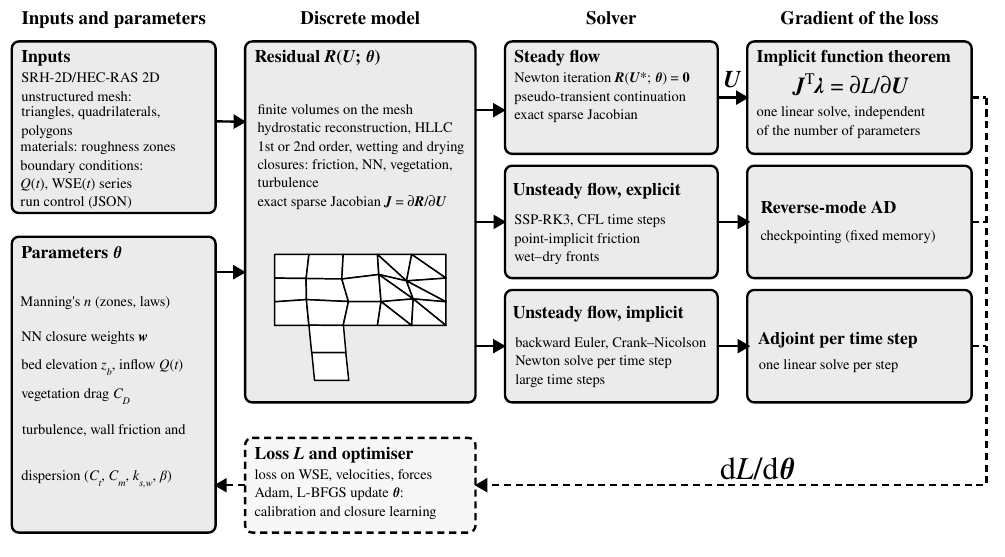}
\caption{Overview of \textit{Hydrograd} v2.0. Inputs (mesh, materials, boundary conditions and run control) define the discrete residual $\mathbf{R}(\mathbf{U};\boldsymbol{\theta})$ on the unstructured mesh. Depending on the problem, the state is computed by steady Newton iteration, explicit time stepping or implicit time stepping, and the gradient of the loss with respect to the parameters $\boldsymbol{\theta}$ is obtained with the matching method: the implicit function theorem, reverse-mode automatic differentiation with checkpointing, or one adjoint linear solve per implicit time step. The gradient drives calibration and closure learning.}
\label{fig:overview}
\end{figure*}

\textit{Hydrograd} v2.0 solves the depth-averaged SWEs with a cell-centred finite-volume method on unstructured meshes. The model is written so that every operation, from the flux computation to the linear solves, can be differentiated with respect to the model parameters. This section describes the equations and closures (Sects.~\ref{sec:equations} and \ref{sec:closures}), the discretisation (Sect.~\ref{sec:discretisation}), the solvers (Sects.~\ref{sec:time} and \ref{sec:jacobian}), the computation of gradients (Sect.~\ref{sec:gradients}), the calibration problem (Sect.~\ref{sec:calibration}) and the software (Sect.~\ref{sec:software}). Figure~\ref{fig:overview} gives an overview.

\subsection{Governing equations}
\label{sec:equations}

The SWEs in conservative form can be written as
\begin{equation}
\frac{\partial \mathbf{U}}{\partial t} + \frac{\partial \mathbf{F}}{\partial x} + \frac{\partial \mathbf{G}}{\partial y} = \mathbf{S},
\label{eq:swe}
\end{equation}
with
\begin{equation}
\mathbf{U} = \begin{pmatrix} \xi \\ q_x \\ q_y \end{pmatrix}, \quad
\mathbf{F} = \begin{pmatrix} q_x \\ h u^2 + \frac{1}{2} g h^2 \\ h u v \end{pmatrix}, \quad
\mathbf{G} = \begin{pmatrix} q_y \\ h u v \\ h v^2 + \frac{1}{2} g h^2 \end{pmatrix},
\label{eq:flux}
\end{equation}
where $\mathbf{U}$ is the state vector,  $\xi$ [m] is the free-surface elevation above the still-water level, $h = \xi + h_s$ [m] is the water depth, $h_s$ [m] is the still-water depth, $q_x = h u$ and $q_y = h v$ [m$^2$\,s$^{-1}$] are the unit discharges, $\mathbf{q} = (q_x, q_y)$ is the unit discharge vector, $u$ and $v$ [m\,s$^{-1}$] are the depth-averaged velocities and $g$ [m\,s$^{-2}$] is the gravitational acceleration. Since $h_s$ does not change in time, $\partial \xi / \partial t = \partial h / \partial t$. The source term is
\begin{equation}
\mathbf{S} = \begin{pmatrix} 0 \\ -g h \dfrac{\partial z_b}{\partial x} - \dfrac{\tau_{bx}}{\rho} + D_x \\[2mm] -g h \dfrac{\partial z_b}{\partial y} - \dfrac{\tau_{by}}{\rho} + D_y \end{pmatrix},
\label{eq:source}
\end{equation}
where $z_b$ [m] is the bed elevation, $\rho$ [kg\,m$^{-3}$] is the water density, $(\tau_{bx}, \tau_{by})$ [N\,m$^{-2}$] is the bed shear stress, $(D_x, D_y)$ [m$^2$\,s$^{-2}$] is the horizontal turbulent diffusion. The bed shear stress and the turbulent diffusion are described in Sect.~\ref{sec:closures}. Bed friction, the bed elevation, the boundary conditions and the closure parameters can all be model parameters $\boldsymbol{\theta}$ with respect to which gradients are computed.

\subsection{Flow-resistance and turbulence closures}
\label{sec:closures}

\subsubsection{Bed friction}

The bed shear stress follows Manning's formula,
\begin{equation}
\frac{\boldsymbol{\tau}_b}{\rho} = \frac{g n^2}{h^{1/3}} |\mathbf{u}| \mathbf{u},
\label{eq:manning}
\end{equation}
where $n$ [s\,m$^{-1/3}$] is Manning's roughness coefficient and $\mathbf{u} = (u, v)$. Several parameterisations of $n$ are available: a constant $n$ per material zone (the usual practice), empirical laws $n(h)$ and $n(h, |\mathbf{u}|, k_s)$ with roughness height $k_s$ [m] \citep{Liu2024NCHRP}, and NN closures. In the NN closures of the USWEs \citep{Liu2025}, a multilayer perceptron predicts either $n$ from local flow variables, e.g. $n = \mathcal{N}(h, |\mathbf{u}|; \mathbf{w})$ with weights $\mathbf{w}$, or the magnitude of the bed shear stress $\boldsymbol{\tau}_b$ directly from $h$ and $|\mathbf{q}|$. The inputs of an NN are scaled to $[-1, 1]$ and the output is mapped into physical bounds with a sigmoid, so that $n$ stays positive and bounded during training. In v2.0, the closure is strictly local, i.e., $n$ in a cell depends only on the variables of that cell. This keeps the Jacobian of the discrete equations sparse (Sect.~\ref{sec:jacobian}) and makes the learned law a transferable function of the flow.

\subsubsection{Horizontal turbulent diffusion}

Horizontal momentum exchange by turbulence is represented by the eddy-viscosity term
\begin{equation}
\mathbf{D} = \nabla \cdot \left(h \nu_t \nabla \mathbf{u}\right),
\label{eq:diffusion}
\end{equation}
where $\nu_t$ [m$^2$\,s$^{-1}$] is the depth-averaged eddy viscosity. Four models are available. (1) A constant $\nu_t$. (2) The parabolic model $\nu_t = C_t u_* h + \nu_0$ used in SRH-2D, with the bed friction velocity $u_* = \sqrt{g}\, n |\mathbf{u}| / (h^{1/6})$ [m\,s$^{-1}$], the coefficient $C_t$ [--] (0.7 by default) and the background molecular viscosity $\nu_0$. (3) A mixing-length model that adds shear-generated mixing,
\begin{equation}
\nu_t = C_t u_* h + (C_m h)^2 |S| + \nu_0, \quad |S| = \sqrt{2 u_x^2 + 2 v_y^2 + (u_y + v_x)^2},
\label{eq:mixing_length}
\end{equation}
where $C_m$ [--] is the mixing-length coefficient, $|S|$ [s$^{-1}$] is the magnitude of the strain rate of the depth-averaged velocity and the subscripts denote partial derivatives. (4) The depth-averaged $k$--$\varepsilon$ model of \citet{Rastogi1978}, which transports the turbulent kinetic energy $k$ [m$^2$\,s$^{-2}$] and its dissipation rate $\varepsilon$ [m$^2$\,s$^{-3}$]:
\begin{align}
\frac{\partial (h k)}{\partial t} + \nabla \cdot (h \mathbf{u} k) &= \nabla \cdot \left(h \frac{\nu_t}{\sigma_k} \nabla k\right) + h\left(P_h + P_{kV} - \varepsilon\right), \label{eq:k}\\
\frac{\partial (h \varepsilon)}{\partial t} + \nabla \cdot (h \mathbf{u} \varepsilon) &= \nabla \cdot \left(h \frac{\nu_t}{\sigma_\varepsilon} \nabla \varepsilon\right) + h\left(c_{1} \frac{\varepsilon}{k} P_h + P_{\varepsilon V} - c_{2} \frac{\varepsilon^2}{k}\right), \label{eq:eps}
\end{align}
with $\nu_t = c_\mu k^2 / \varepsilon$, the shear production $P_h = \nu_t |S|^2$, and the production by bed shear
\begin{equation}
P_{kV} = \frac{c_f |\mathbf{u}|^3}{h}, \quad P_{\varepsilon V} = 3.6\, c_2 \sqrt{c_\mu}\, \frac{c_f^{5/4} |\mathbf{u}|^4}{h^2}, \quad c_f = \frac{u_*^2}{|\mathbf{u}|^2} = \frac{g n^2}{h^{1/3}},
\label{eq:ke_sources}
\end{equation}
where $c_\mu = 0.09$, $c_1 = 1.44$, $c_2 = 1.92$, $\sigma_k = 1.0$ and $\sigma_\varepsilon = 1.3$ are the standard constants and $c_f$ [--] is the bed friction coefficient. In uniform flow, Eqs.~(\ref{eq:k})--(\ref{eq:ke_sources}) give $\nu_t = u_* h / 12.96 \approx 0.077\, u_* h$, the classic result of \citet{Rastogi1978}. With the $k$--$\varepsilon$ model, the state vector is extended to $\mathbf{U} = (\xi, q_x, q_y, h k, h \varepsilon)$.

\subsection{Unstructured mesh and spatial discretisation}
\label{sec:discretisation}

\subsubsection{Mesh representation}

\begin{figure*}[htp]
\includegraphics[width=17cm]{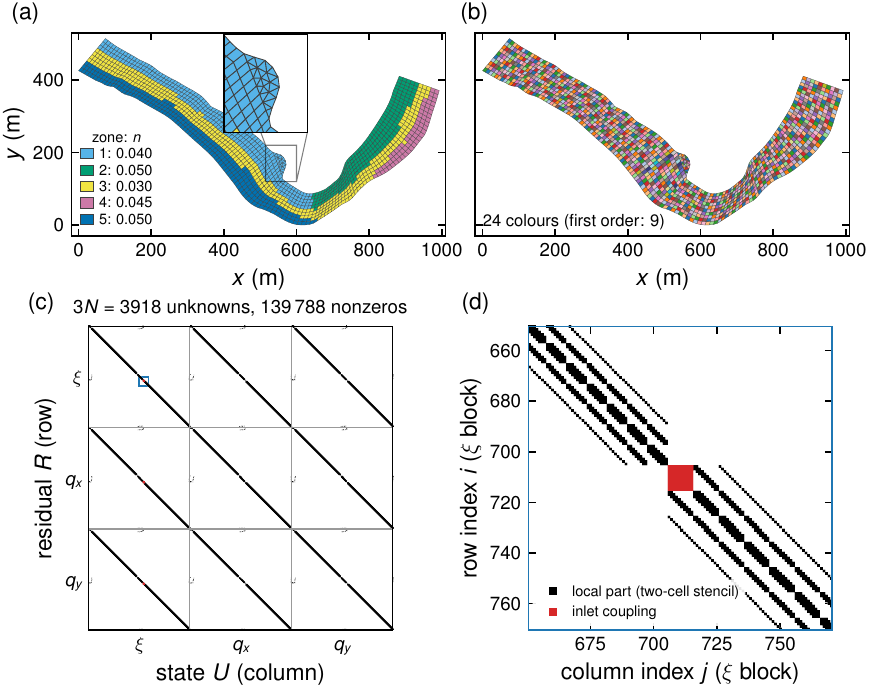}
\caption{Unstructured mesh and exact sparse Jacobian of the Savannah River case: (a) SRH-2D mesh of 1,306 cells coloured by Manning roughness zone, with a close-up of the transition from quadrilateral to triangular cells; (b) distance-2 colouring of the cells used to compute the Jacobian (24 colours at second order, 9 at first order); (c) sparsity pattern of $\partial \mathbf{R}/\partial \mathbf{U}$, arranged in blocks of the state variables $\xi$, $q_x$ and $q_y$; (d) close-up of the $\xi$--$\xi$ block near the inlet, showing the two-cell-wide stencil of the second-order scheme and the dense coupling among the inlet cells (red).}
\label{fig:mesh}
\end{figure*}

\textit{Hydrograd} reads the meshes, materials and boundary conditions of SRH-2D (\texttt{.srhgeom}, \texttt{.srhmat} and \texttt{.srhhydro} files), which can be created with the Surface-water Modeling System (SMS) and other tools used by practitioners. Meshes built for HEC-RAS 2D \citep{Brunner2016} can be used as well: the python package \textit{pyHMT2D} \citep{Liu2026pyHMT2D} reads the mesh and the cell values of Manning's $n$ of a HEC-RAS 2D plan from its HDF5 file and writes the \texttt{.srhgeom} and \texttt{.srhmat} files, so that only the boundary conditions in the \texttt{.srhhydro} file remain to be specified. Cells may be triangles, quadrilaterals or general polygons (Fig.~\ref{fig:mesh}). From the mesh, the model builds the lists of interior faces (each with an owner cell, a neighbour cell, a unit normal pointing from the owner to the neighbour, and a length) and of boundary faces (each with its cell, outward normal, length and boundary type), together with cell areas and centroids. The residual of the finite-volume method is computed in three steps that are the same on any mesh: the states of the two cells of every face are gathered, the flux through every face is computed, and the fluxes are added to the two cells with opposite signs. Written this way, the computation runs efficiently on CPUs and GPUs and can be differentiated by AD.

\subsubsection{Fluxes and well-balanced bed treatment}

The flux through a face is computed with the HLLC approximate Riemann solver \citep{Toro2001} in the frame of the face normal, with the two-rarefaction wave-speed estimates of \citet{Toro2001}, which are exact for dry states. The bed slope is treated with the hydrostatic reconstruction of \citet{Audusse2004}: at a face between cells $L$ and $R$, the depths are reconstructed above the higher bed, $z_f = \max(z_L, z_R)$,
\begin{equation}
h_L^* = \max(0, h_L + z_L - z_f), \quad h_R^* = \max(0, h_R + z_R - z_f),
\label{eq:hr}
\end{equation}
the Riemann problem is solved between the reconstructed states, and the pressure corrections $\frac{1}{2} g (h_L^2 - h_L^{*2})$ and $\frac{1}{2} g (h_R^2 - h_R^{*2})$ replace the bed-slope source. The scheme is exactly well balanced for a lake at rest, including partially wet cells and emerged bed, preserves non-negative depths under a time-step restriction, and conserves mass. To keep velocities bounded in nearly dry cells without clipping the depth, the velocity is computed as $u = \sqrt{2}\, h q_x / \sqrt{h^4 + \max(h^4, \epsilon^4)}$ with $\epsilon = 10^{-3}$\,m \citep{Kurganov2007}, which equals $q_x / h$ in wet cells.

\subsubsection{Second-order reconstruction}

At second order, the free-surface elevation $h + z_b$, the depth and the velocities are reconstructed linearly to the faces from Green--Gauss cell gradients, limited with the limiter of \citet{Venkatakrishnan1995} (default) or of \citet{Barth1989}. The centred bed source of \citet{Audusse2005} keeps the scheme well balanced. Reconstructing the free surface rather than the bed keeps a lake at rest exactly flat. Next to dry cells, the reconstruction is blended continuously between second and first order: the limited slope of a cell is multiplied by a weight that rises linearly from zero (first order) at $h_{\min} = \epsilon$ to one (second order) at $h_{\min} = 10\epsilon$, where $h_{\min}$ is the smallest depth among the cell and its neighbours. With the blend, the fluxes vary continuously with the cell depths instead of jumping when a cell switches between first and second order, which Newton's method needs in the implicit and steady solvers.

\subsubsection{Boundary conditions}

Inflow boundaries prescribe the discharge $Q$ [m$^3$\,s$^{-1}$], which is distributed over the wet boundary faces in proportion to their conveyance $L_f h^{5/3}/n$, where $L_f$ is the face length. Outflow boundaries prescribe the water surface elevation (WSE). Walls reflect the discharge and symmetry boundaries reflect its normal component. $Q$ and the WSE can be time series and can themselves be model parameters, which allows inflow hydrographs to be inverted (Sect.~\ref{sec:hydrograph}).

\subsection{Time integration and steady states}
\label{sec:time}

The discrete equations of all cells form the semi-discrete system
\begin{equation}
\frac{\mathrm{d}\mathbf{U}}{\mathrm{d}t} = \mathbf{R}(\mathbf{U}; \boldsymbol{\theta}, t),
\label{eq:semi}
\end{equation}
where $\mathbf{U}$ now collects the cell values of all state variables and $\mathbf{R}$ is the residual, which is the net flux into each cell divided by its area, plus the source terms. This semi-discrete system is integrated in time with either explicit or implicit methods, or solved for a steady state.

\subsubsection{Explicit time stepping}

For unsteady flows with wetting and drying, Eq.~(\ref{eq:semi}) is integrated with the three-stage strong-stability-preserving Runge--Kutta (SSP-RK3) method \citep{Shu1988,Gottlieb2001}. The time step is set by the Courant number (0.5 at first order and 0.25 at second order), the smallest cell size and the largest wave speed. With the mixing-length and $k$--$\varepsilon$ models, it also satisfies the explicit limit of the turbulent diffusion. The step is recomputed from the state at every step and is treated as a constant in the gradient, so the gradient is exact for the discrete scheme with the realised step sizes but omits their dependence on the parameters. The number of steps therefore varies with the parameters, and the loss function of a calibration is only piecewise smooth. Within each stage, the sink terms that are proportional to their own state variable are treated point-implicitly. Bed friction can be written as $-c\,\mathbf{q}$ with $c = g n^2 |\mathbf{q}| / h_\epsilon^{7/3}$ [s$^{-1}$], where $h_\epsilon = \sqrt{h^2 + \epsilon^2}$ regularises the depth in nearly dry cells, and the dissipation terms of $h k$ and $h \varepsilon$ as $-(\varepsilon/k)\, h k$ and $-c_2 (\varepsilon/k)\, h \varepsilon$. All other terms are first advanced explicitly to an intermediate state $\mathbf{U}^* = \mathbf{U}^n + \Delta t\, \mathbf{R}_e(\mathbf{U}^n)$, where $\mathbf{U}^n$ denotes the state at the start of the stage. The sink is then applied implicitly with its rate $c$ also evaluated at $\mathbf{U}^n$, which gives $\mathbf{U}^{n+1} = \mathbf{U}^* / (1 + \Delta t\, c)$ for each affected component. Unlike the explicit form $\mathbf{U}^*(1 - \Delta t\, c)$, this can only reduce the magnitude of the variable and never changes its sign, so friction cannot reverse the flow even in very thin water films where $c$ is large and explicit friction would be unstable. The general-purpose adaptive Runge--Kutta methods of the diffrax library \citep{Kidger2021} are also available.

\subsubsection{Implicit time stepping}

For long, slowly varying events or meshes with very small cells, the $\vartheta$-scheme
\begin{equation}
\mathbf{G}(\mathbf{U}^{k+1}) = \mathbf{U}^{k+1} - \mathbf{U}^{k} - \Delta t \left[\vartheta \mathbf{R}(\mathbf{U}^{k+1}, t^{k+1}) + (1 - \vartheta) \mathbf{R}(\mathbf{U}^{k}, t^{k})\right] = \mathbf{0}
\label{eq:theta}
\end{equation}
is solved at every time step with Newton's method, where $\vartheta = 1$ gives the backward Euler method and $\vartheta = 1/2$ the Crank--Nicolson method. Each Newton iteration solves a sparse linear system with the exact Jacobian $\partial \mathbf{G} / \partial \mathbf{U} = \mathbf{I} - \vartheta \Delta t \mathbf{J}$, where $\mathbf{J} = \partial \mathbf{R} / \partial \mathbf{U}$ (Sect.~\ref{sec:jacobian}), with a damped update. The Newton direction $\Delta\mathbf{U}$ is scaled by the step fraction $\alpha \in \{1, 1/2, 1/4, 1/8\}$ that gives the smallest residual $\max |\mathbf{G}(\mathbf{U} + \alpha \Delta\mathbf{U})|$. All four candidates are evaluated at once, which keeps the control flow fixed for compilation, and a candidate with a non-finite residual is discarded. The Jacobian is reused over successive iterations and re-formed only when the residual has fallen by less than a fixed ratio. The time step is fixed rather than adaptive, so that the sequence of steps does not change with the parameters. This removes the main source of non-smoothness of the calibration loss of the explicit solver; the wet/dry treatment and the limiter still make the discrete loss only piecewise smooth (Sect.~\ref{sec:limitations}).

\subsubsection{Steady states}

A steady state is a root $\mathbf{U}^*$ of $\mathbf{R}(\mathbf{U}^*; \boldsymbol{\theta}) = \mathbf{0}$. It is found by pseudo-transient continuation. Each iteration solves
\begin{equation}
\left(\frac{\mathbf{I}}{\Delta\tau_k} - \mathbf{J}\right) \Delta \mathbf{U} = \mathbf{R}(\mathbf{U}_k), \quad \mathbf{U}_{k+1} = \mathbf{U}_k + \Delta \mathbf{U},
\label{eq:ptc}
\end{equation}
with the pseudo-time step $\Delta\tau_k = \min[f_k \Delta\tau_0 (\|\mathbf{R}_0\|_\infty / \|\mathbf{R}_k\|_\infty)^p, \Delta\tau_{\max}]$ of switched evolution relaxation \citep{Mulder1985}. Here $\Delta\tau_0$ [s] is a user-specified initial pseudo-time step (1\,s by default, set per case), $\mathbf{R}_0$ is the residual of the initial guess, $\|\cdot\|_\infty$ is the maximum over all cells and components, $p = 1.5$ is the growth exponent and $\Delta\tau_{\max}$ is an upper bound ($10^{12}$\,s by default, so that it is inactive). Far from the solution, Eq.~(\ref{eq:ptc}) behaves like an implicit time step and is robust from a poor initial guess. As the residual falls, $\Delta\tau_k$ grows and the iteration becomes Newton's method with quadratic convergence. After each update, the depth is clipped at zero and the Jacobian is re-formed at every iteration. A step whose residual is not finite or grows by more than a factor of 10 is rejected. In such cases, the state is kept and the safeguard factor $f_k$, initially 1, is divided by 10. After each accepted step, $f_k$ is doubled, up to 1. The iteration stops when $\|\mathbf{R}_k\|_\infty < 10^{-10}$. Steady states are reached to residuals near machine precision in a few iterations for subcritical river flow (7 for the Savannah River case) and in about 100 for a transcritical flow with a hydraulic jump (Sect.~\ref{sec:verification}).

\subsection{Exact sparse Jacobian on unstructured meshes}
\label{sec:jacobian}

The implicit and steady solvers and the steady gradients (Sect.~\ref{sec:gradients}) need the Jacobian $\mathbf{J} = \partial \mathbf{R} / \partial \mathbf{U}$. On a mesh with $N$ cells and $n_v$ state variables per cell, $\mathbf{J}$ has $(n_v N)^2$ entries, but the residual of a cell depends only on the cell itself and its neighbours (first order), or on its neighbours' neighbours (second order and gradient-based closures). \textit{Hydrograd} computes the nonzero entries of $\mathbf{J}$ exactly, without finite differences, with the compression technique of \citet{Curtis1974}. The cells are coloured such that no two cells of the same colour influence a common cell (a distance-2 colouring of the mesh graph). 

A column of $\mathbf{J}$ is the derivative of $\mathbf{R}$ with respect to one state variable in one cell, and forward-mode AD returns it as one Jacobian-vector product (JVP). Computing the columns one at a time would need $n_v N$ JVPs. Because cells of the same colour never affect a common cell, their columns have no nonzero entries in the same row. As a result, they can be computed in parallel: one JVP seeded with ones in variable $j$ of every cell of one colour returns all of those columns at once. The number of JVPs is thus $n_v$ times the number of colours. This number depends on the mesh connectivity and the width of the stencil but not on the mesh size. The Savannah River mesh of 1,306 cells needs 9 colours at first order and 24 at second order, i.e. 27 and 72 JVPs instead of 3,918 (Fig.~\ref{fig:mesh}).

Inflow boundaries break this locality. The prescribed discharge is distributed over the faces of an inlet in proportion to their conveyance, so the inflow through one inlet cell depends on the depths of all cells along that inlet. Every inlet cell therefore influences every other, and the colouring above would need as many colours as there are inlet cells. Instead, the Jacobian is split into a local part, which is computed by colouring as described, and a small correction for each inlet that carries the coupling among its cells \citep{Hager1989}.

\subsection{Gradients}
\label{sec:gradients}

Calibration and closure learning minimise a loss $\mathcal{L}(\mathbf{U}; \boldsymbol{\theta})$ that measures the misfit between the simulated and observed quantities (Sect.~\ref{sec:calibration}). The gradient $\mathrm{d}\mathcal{L} / \mathrm{d}\boldsymbol{\theta}$ is computed exactly, i.e. to the precision of the discrete model, with one of three methods chosen according to the problem.

\subsubsection{Steady flows: implicit function theorem}

At a steady state, $\mathbf{R}(\mathbf{U}^*(\boldsymbol{\theta}); \boldsymbol{\theta}) = \mathbf{0}$ for all $\boldsymbol{\theta}$. Differentiating this identity gives $\mathbf{J}\, \mathrm{d}\mathbf{U}^* / \mathrm{d}\boldsymbol{\theta} = -\partial \mathbf{R} / \partial \boldsymbol{\theta}$ (the implicit function theorem), and the gradient of the loss follows from one linear solve with the transposed Jacobian,
\begin{equation}
\mathbf{J}^{\mathrm{T}} \boldsymbol{\lambda} = \frac{\partial \mathcal{L}}{\partial \mathbf{U}}, \quad
\frac{\mathrm{d}\mathcal{L}}{\mathrm{d}\boldsymbol{\theta}} = \frac{\partial \mathcal{L}}{\partial \boldsymbol{\theta}} - \boldsymbol{\lambda}^{\mathrm{T}} \frac{\partial \mathbf{R}}{\partial \boldsymbol{\theta}},
\label{eq:ift}
\end{equation}
where $\boldsymbol{\lambda}$ is the adjoint state and $\boldsymbol{\lambda}^{\mathrm{T}} \partial \mathbf{R} / \partial \boldsymbol{\theta}$ is one reverse-mode AD product. The cost is one linear solve, which is independent of the number of parameters. The gradient depends only on the steady state, not on how the solver reached it. Differentiating through the Newton iterations or through a time march to steady state, as is done in \textit{Hydrograd} v1.0, would give a gradient that depends on the computational path and on the number of iterations (Sect.~\ref{sec:steady_ude}).

\subsubsection{Unsteady flows with explicit time stepping}

For explicit time stepping, the gradient is computed by reverse-mode AD through the time steps, which is equivalent to integrating the discrete adjoint equations backwards in time. Storing every intermediate state would require memory proportional to the number of time steps. Instead, only a fixed number of checkpoints (64 by default, or as many as fit a given memory budget) are stored, and the states between checkpoints are recomputed during the backward pass \citep{Griewank2000}. The gradient then costs about 4 to 8 forward runs, with memory independent of the simulation length. As noted in Sect.~\ref{sec:time}, the state-dependent step sizes are held fixed in this gradient, and the loss is only piecewise smooth in the parameters.

\subsubsection{Unsteady flows with implicit time stepping}

For the implicit scheme of Eq.~(\ref{eq:theta}), each time step is itself a root-finding problem, $\mathbf{G}(\mathbf{U}^{k+1}; \mathbf{U}^k, \boldsymbol{\theta}) = \mathbf{0}$. Its gradient follows from the implicit function theorem as in Eq.~(\ref{eq:ift}), with one transposed linear solve per time step and no differentiation of the Newton iterations. With a fixed $\Delta t$, the loss has none of the step-count jumps of the explicit solver, and a gradient costs about 2 to 3 forward runs for the cases in this work (Table~\ref{tab:gradients}).

\subsubsection{Forward-mode sensitivities}

When the number of parameters is small and the sensitivities of many outputs are wanted, forward-mode AD gives $\mathrm{d}\mathbf{U}/\mathrm{d}\boldsymbol{\theta}$ directly through the explicit solvers (Sect.~\ref{sec:gradient_checks}). The implicit solver provides reverse-mode derivatives only, so its sensitivities to a few parameters are obtained by finite differences, which can be checked against the adjoint (Sect.~\ref{sec:savannah_unsteady}). For steady states, $\mathrm{d}\mathbf{U}^*/\mathrm{d}\boldsymbol{\theta} = -\mathbf{J}^{-1} \partial \mathbf{R}/\partial \boldsymbol{\theta}$ is obtained by factorising $\mathbf{J}$ once and solving for the columns of $\partial \mathbf{R}/\partial \boldsymbol{\theta}$ by back-substitution, one per parameter; the inverse $\mathbf{J}^{-1}$ itself is never formed.

\subsection{Calibration and closure learning}
\label{sec:calibration}

For calibration and closure learning, the loss function combines normalised misfits of the observed quantities,
\begin{equation}
\mathcal{L} = w_{\mathrm{WSE}} \mathcal{L}_{\mathrm{WSE}} + w_{\mathbf{u}} \mathcal{L}_{\mathbf{u}} + w_{\mathrm{bound}} \mathcal{L}_{\mathrm{bound}},
\label{eq:loss}
\end{equation}
where $\mathcal{L}_{\mathrm{WSE}}$ and $\mathcal{L}_{\mathbf{u}}$ are the mean squared misfits of the WSE and the velocity at the observation cells and times, normalised by the range of the observed WSE and the largest observed velocity, $\mathcal{L}_{\mathrm{bound}}$ penalises parameters outside physical bounds and the $w$ are weights. Parameters that must stay positive, such as Manning's $n$, can be optimised in logarithmic form. The optimisers of the Optax library \citep{Optax2020}, e.g. Adam \citep{Kingma2015} and the limited-memory BFGS method with a line search \citep{Liu1989}, are used.

\subsection{Software}
\label{sec:software}

\textit{Hydrograd} v2.0 is a Python package (Python 3.11 or newer) built on JAX \citep{Bradbury2018}, with diffrax \citep{Kidger2021}, Optax \citep{Optax2020}, NumPy and SciPy. Double precision is used as default. A case is defined by the SRH-2D input files and a run-control file in JSON format that is compatible with v1.0, and is run with the command \texttt{hydrograd run}. The run-control file selects one of four modes: forward simulation, sensitivity analysis, inversion and closure learning (USWEs). It also sets the numerical options (scheme, spatial order, limiter, solver, checkpoints). Results are written in VTK format for visualisation in ParaView and in JSON format. The model has been tested on Linux CPUs and NVIDIA GPUs and runs wherever JAX is supported. The repository contains a user manual, the examples of this paper and the scripts that reproduce every figure. The code is distributed under the MIT licence (see the code and data availability section). The differences between v2.0 and v1.0, including bugs of v1.0 that were found and corrected during the port, are listed in Appendix~\ref{app:differences}.

\section{Verifications}
\label{sec:verification}

This section verifies that the SWEs are solved correctly. The comparison with another model on a real river is part of the evaluation in Sect.~\ref{sec:applications}.

\subsection{Verification against analytical solutions}

Three analytical solutions of the SWEs are used: the dam break over a wet bed \citep{Stoker1992}, the dam break over a dry bed (Ritter's solution; \citealt{Delestre2013}) and the transcritical flow over a bump with a hydraulic jump \citep{Goutal1997}. The lake at rest over an emergent island tests the well-balanced property of the scheme. 

\begin{figure*}[htp]
\includegraphics[width=17cm]{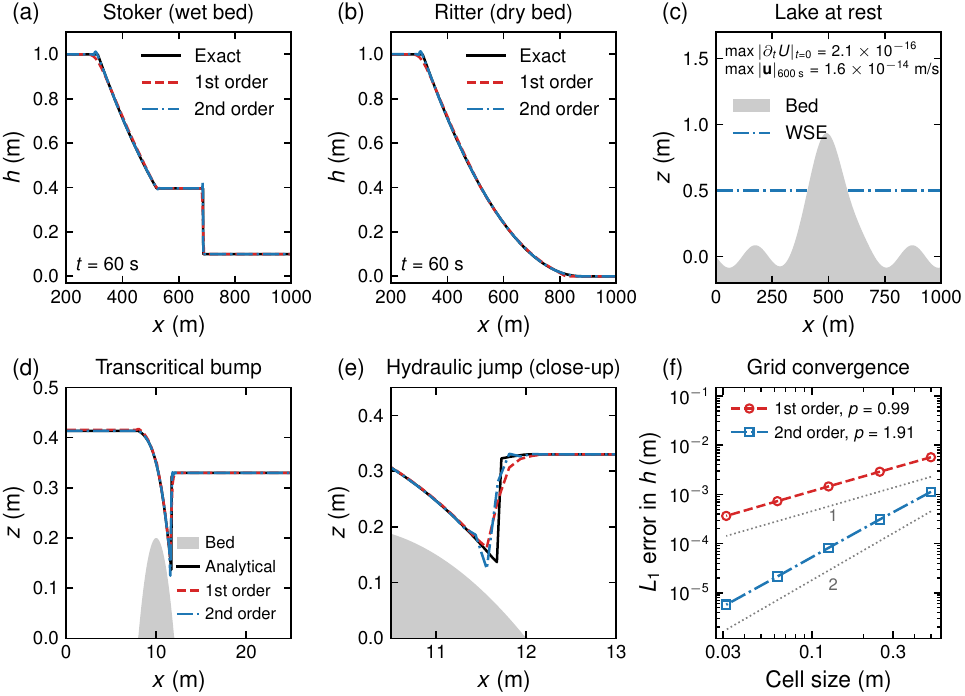}
\caption{Verification against analytical solutions: (a) dam break over a wet bed \citep{Stoker1992}; (b) dam break over a dry bed (Ritter's solution); (c) lake at rest over an emergent island; (d) transcritical flow over a bump with a hydraulic jump; (e) close-up of the hydraulic jump in (d); (f) error against mesh size, with the observed orders of accuracy.}
\label{fig:verification}
\end{figure*}

Figure~\ref{fig:verification} compares \textit{Hydrograd} v2.0 with analytical solutions at first and second order. For the dam break over a wet bed \citep{Stoker1992}, the $L_1$ error of the depth is 2.34\,mm at first order and 0.47\,mm at second order, and the $L_1$ error of the velocity is 0.011 and 0.0025\,m\,s$^{-1}$. For the dam break over a dry bed, the $L_1$ errors of the depth are 2.61 and 0.39\,mm for the first and second order, respectively. In both dam break cases, the depth stays non-negative, and the relative change of the total water volume is below $10^{-14}$.

For a lake at rest over an island that emerges from the water surface, the residual of the initial state is $2.1 \times 10^{-16}$. After 600\,s of simulation, the largest velocity is $1.6 \times 10^{-14}$\,m\,s$^{-1}$. This confirms that the hydrostatic reconstruction keeps the scheme well balanced with wet and dry cells, at both orders.

For the transcritical flow over a bump in a frictionless channel, with a hydraulic jump downstream of the crest, the root-mean-square (RMS) error of the WSE against the analytical solution is 1.5\,mm at first order and 0.07\,mm at second order away from the jump (more than 0.5\,m from it). The largest error is in the single cell that contains the jump, where any shock-capturing scheme smears the discontinuity. On a sequence of meshes from 50 to 800 cells, the observed order of accuracy on a smooth subcritical flow is 0.99 at first order and 1.91 at second order (Fig.~\ref{fig:verification}f), as expected from the design of the scheme.

\subsection{Verification of the steady solver}

The steady solver was compared with long unsteady simulations to the same steady state. For the transcritical flow over a bump case in the previous section, with water initially at rest with the downstream level of 0.33 m, the WSE difference between the steady solver and the long unsteady simulation after 2,000\,s of time marching with a relative tolerance of $10^{-8}$ is $2.0 \times 10^{-9}$\,m. For the Savannah River case (Sect.~\ref{sec:savannah_srh2d}), using a pre-simulated SRH-2D solution as starting point, the WSE difference after 3,000\,s is $7.5 \times 10^{-7}$\,m, which is consistent with the residual of $1.4 \times 10^{-7}$ that the unsteady run still had. The Newton solve reached a residual of $10^{-13}$ in 7 iterations for the Savannah River case. For the bump case, whose hydraulic jump (discontinuity) makes the steady problem stiffer, Newton's method needed about 100 iterations. 

\section{Evaluation and applications}
\label{sec:applications}

This section evaluates the model against SRH-2D on a real river reach (Sect.~\ref{sec:savannah_srh2d}) and demonstrates its new capabilities: steady roughness calibration (Sect.~\ref{sec:steady_inversion}), steady closure learning (Sect.~\ref{sec:steady_ude}), the inversion of an inflow hydrograph (Sect.~\ref{sec:hydrograph}) and unsteady roughness calibration (Sect.~\ref{sec:savannah_unsteady}). Section~\ref{sec:performance} reports the computational performance. Unless stated otherwise, the second-order scheme was used, and the timings were measured on one node of the Roar Collab cluster of Penn State Institute for Computational and Data Sciences with 8 cores of an Intel Xeon Gold 6342 CPU (2.8\,GHz). Some cases were run on an NVIDIA A100 GPU.

\subsection{Savannah River: comparison with SRH-2D}
\label{sec:savannah_srh2d}

The Savannah River case of \textit{Hydrograd} v1.0 \citep{Liu2025} is a 1\,km reach of a river bend (Fig.~\ref{fig:mesh}a), meshed in SMS for SRH-2D with 1,306 cells (1,280 quadrilaterals and 26 triangles), with sizes from 5.4 to 12.1\,m. Five material zones represent the main channel and the overbanks, with Manning's $n$ from 0.030 to 0.050\,s\,m$^{-1/3}$ (Fig.~\ref{fig:mesh}a). The inflow is 187\,m$^3$\,s$^{-1}$, and the WSE at the outlet is prescribed. Figure~\ref{fig:mesh} also shows the colouring of the mesh and the sparsity of the Jacobian. At second order, 24 colours and 3 unknowns ($\xi$, $hu$, and $hv$), i.e. 72 (=$24 \times 3$) forward-mode products, recover all 139,788 nonzero entries of the Jacobian for the 3,918 unknowns. 

\begin{figure*}[htp]
\includegraphics[width=17cm]{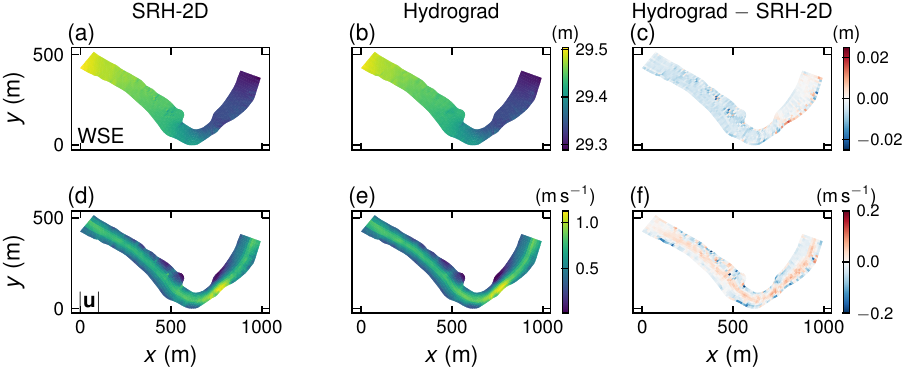}
\caption{Steady flow in the Savannah River comparison between SRH-2D (left column) and \textit{Hydrograd} v2.0 (middle column), and their difference (right column): (a)--(c) water-surface elevation (WSE); (d)--(f) flow velocity magnitude $|\mathbf{u}|$.}
\label{fig:savannah}
\end{figure*}

The steady solution of \textit{Hydrograd} v2.0 was compared with that of SRH-2D. Starting from a pre-simulated SRH-2D solution as the initial guess, Newton's method in \textit{Hydrograd} converged in 7 iterations (1.3\,s) to a residual of $10^{-13}$. The root mean square (RMS) difference of the WSE between the two models is 4.7\,mm, with a maximum of 20\,mm, and the RMS difference of the flow velocity is 0.037\,m\,s$^{-1}$ (Fig.~\ref{fig:savannah}). The differences are largest along the banks and in the bend, where the two models resolve the velocity gradients differently. They are small compared with the WSE drop of about 0.2\,m through the reach. Without eddy viscosity the differences increase to 6.3\,mm and 0.053\,m\,s$^{-1}$, and with the first-order scheme to 27.5\,mm and 0.054\,m\,s$^{-1}$. The two models therefore agree within the differences expected between independent 2D models, provided they include the same physics and use comparable numerical methods.

\subsection{Steady roughness calibration}
\label{sec:steady_inversion}

\begin{figure*}[htp]
\includegraphics[width=17cm]{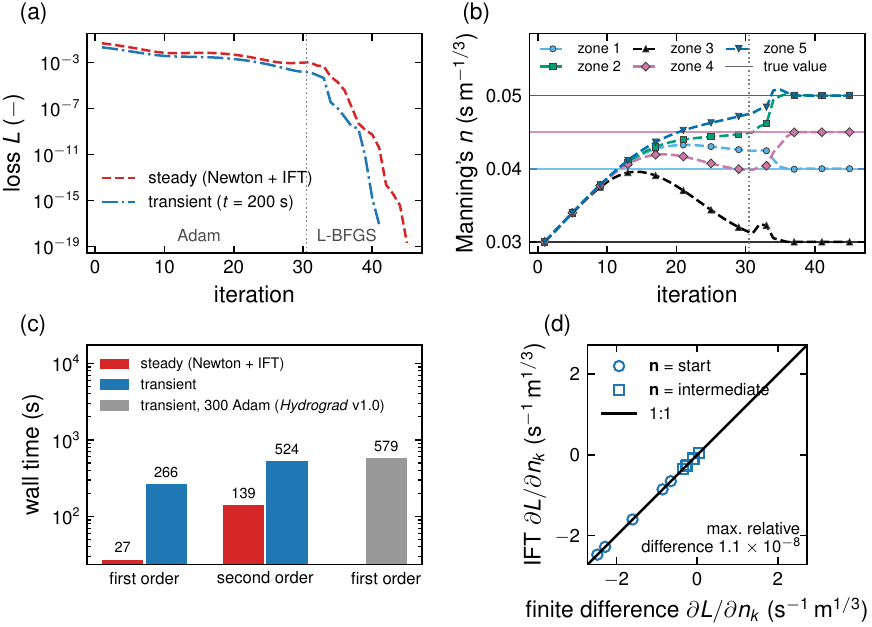}
\caption{Steady calibration of Manning's $n$ in five zones of the Savannah River: (a) loss against iteration for the steady approach (Newton iteration and gradients from the implicit function theorem, IFT) and the transient approach (time marching for 200\,s and reverse-mode AD), with Adam for 30 iterations followed by L-BFGS; (b) Manning's $n$ of each zone against iteration (steady approach), with the true values; (c) wall time of the whole inversion at first and second order, and with the settings of the \textit{Hydrograd} v1.0 example; (d) gradient from the IFT against central finite differences at the start and at an intermediate point of the inversion.}
\label{fig:steady_inversion}
\end{figure*}

To demonstrate steady calibration, Manning's $n$ of the five zones shown in Fig.~\ref{fig:mesh}a was inverted from the steady WSE and velocity fields computed with the true values $n = (0.040, 0.050, 0.030, 0.045, 0.050)$\,s\,m$^{-1/3}$, starting from $n = 0.03$\,s\,m$^{-1/3}$ in all zones. The optimisation used 30 iterations of Adam followed by L-BFGS. With steady states and the gradients of the implicit function theorem, the inversion recovered all five values to within $1.1 \times 10^{-10}$ in 44 iterations (Fig.~\ref{fig:steady_inversion}a, b). The gradients agree with finite differences to $1.1 \times 10^{-8}$ (Fig.~\ref{fig:steady_inversion}d).

For comparison, the same inversion was run with the approach of \textit{Hydrograd} v1.0: each loss evaluation marched the unsteady equations for 200\,s, with gradients by reverse-mode AD through the time steps. A steady gradient cost 1.5\,s, against 8.3\,s for a gradient through the time march with reverse model AD (18.6\,s with forward mode AD). The whole inversion took 139\,s with steady states, against 524\,s with time marching, a factor of 3.8 (Fig.~\ref{fig:steady_inversion}c). These totals include the compilation in the first iteration. At first order, the factor is 9.8 (27 against 266\,s). The smaller gain at second order reflects the higher cost of the Jacobian, which has 2.5 times the nonzeros of the first-order one and needs 72 instead of 27 Jacobian-vector products. For this small mesh, the Jacobian accounts for most of the 0.25\,s of a Newton iteration. With the settings of the \textit{Hydrograd} v1.0 example (300 Adam iterations, first order), the time-marching inversion took 579\,s.

\subsection{Steady closure learning}
\label{sec:steady_ude}

The USWEs embed an NN closure for flow resistance in the SWEs \citep{Liu2025}. For the transcritical flow over a bump case, instead of a constant Manning's $n$, \citet{Liu2025} introduced a depth-dependent Manning's $n$ in the form of a sigmoid function as
\begin{equation}\label{eqn:sigmoid_n_h}
    n(h) = n_{\text{lower}} + \frac{ n_{\text{upper}} - n_{\text{lower}} }{1 + e^{k(h-h_{\text{mid}})}}   
\end{equation}
where $n_{\text{lower}}$ and $n_{\text{upper}}$ are the lower and upper bounds of the Manning's $n$ values for deep and shallow flow depth, respectively, and $k$ [1/m] and $h_{\text{mid}}$ [m] are the parameters that control the shape and the transition point of the function. In this example, Manning's $n$ value drops from $n_{\text{upper}}$ = 0.06 to $n_{\text{lower}}$ = 0.03\,s\,m$^{-1/3}$ around $h_{\text{mid}} = 0.3$\,m, with $k = 100$\,m$^{-1}$.

An NN $n = \mathcal{N}(h)$ with two hidden layers of three neurons (22 weights) learns this law from the steady WSE and velocities along the channel (200 cells). We trained the same NN, from the same initial weights and with 100 iterations of Adam, in two ways. The first was the approach of time marching for 100\,s with an adaptive Runge--Kutta method and a tight tolerance, with gradients by reverse-mode AD with checkpointing. The second was the steady approach: Newton iteration and gradients from the implicit function theorem.

\begin{figure*}[htp]
\includegraphics[width=17cm]{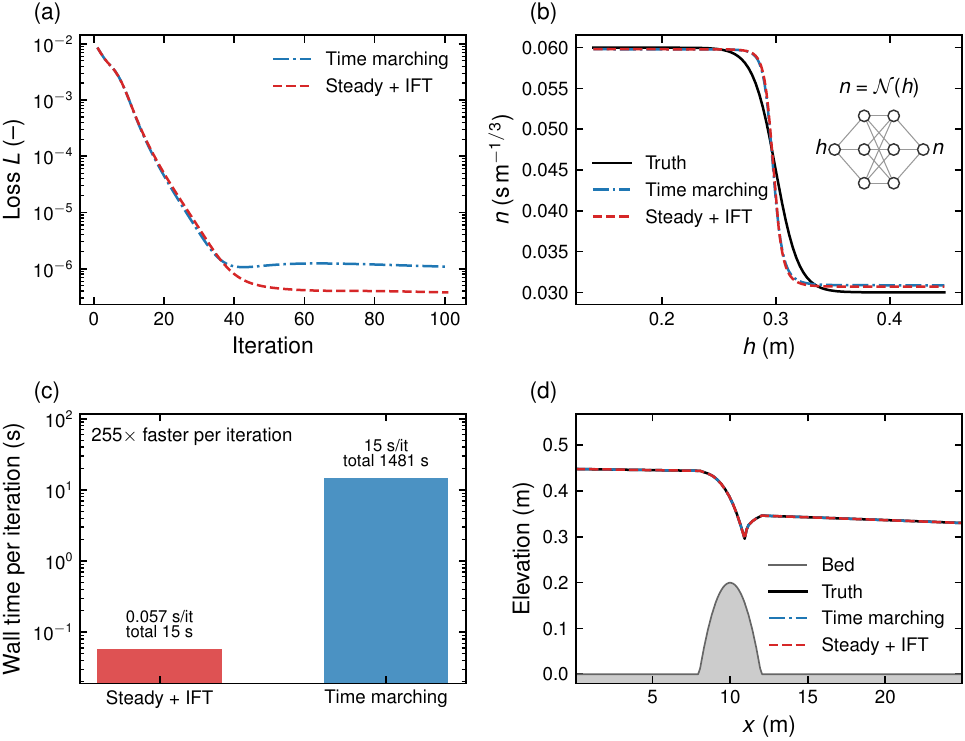}
\caption{Learning an NN closure $n = \mathcal{N}(h)$ in the bump case of \citet{Liu2025} with time marching and with steady states: (a) loss against Adam iteration; (b) learned $n(h)$ and the true law over the range of depths in the data; (c) wall time per iteration and in total; (d) WSE along the channel computed with the true law and with the two learned closures.}
\label{fig:steady_ude}
\end{figure*}

Both approaches learned the same closure (Fig.~\ref{fig:steady_ude}b), with RMS errors of $n(h)$ of $1.6 \times 10^{-3}$\,s\,m$^{-1/3}$ over the range of depths in the data. The largest errors are at the steep transition near $h = 0.3$\,m, where few cells provide data. The steady approach took 0.057\,s per iteration and 15\,s in total, against 14.5\,s per iteration and 1,481\,s for time marching. The steady plus IFT approch is 255 times faster per iteration and 96 times faster in total (Fig.~\ref{fig:steady_ude}c). The steady approach also reached a lower loss ($3.8 \times 10^{-7}$ against $1.1 \times 10^{-6}$; Fig.~\ref{fig:steady_ude}a). After 100\,s of time marching, the flow was not fully steady. The WSE still changed at $3.7 \times 10^{-5}$\,m\,s$^{-1}$ and differed from the steady state by up to 0.23\,mm. This remaining transient set a floor to the loss of the time-marching approach after about 40 iterations. The gradient of the time-marching approach is the gradient of the solution at $t = 100$\,s, not of the steady state, and its error depends on how far the run is from steady state. The steady approach removes this dependence, as discussed in Sect.~\ref{sec:gradients}.

\subsection{Inversion of an inflow hydrograph}
\label{sec:hydrograph}

\begin{figure*}[htp]
\includegraphics[width=17cm]{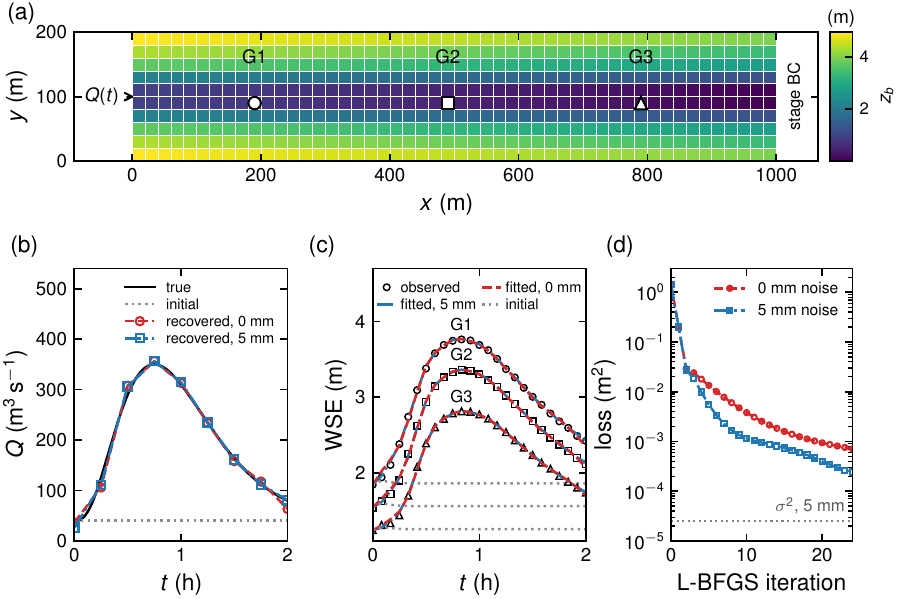}
\caption{Inversion of an inflow hydrograph from stage records: (a) bed elevation of the compound valley, with the three gauges G1--G3, the inflow and the stage boundary; (b) true, initial and recovered hydrographs without noise and with 5\,mm noise; (c) observed and fitted WSE at the gauges; (d) loss against L-BFGS iteration, with the variance of the 5\,mm noise.}
\label{fig:hydrograph}
\end{figure*}

Inflow hydrographs are often unknown for ungauged tributaries or historical floods, while stage records may exist. Even for gauged sites, direct measurement of flow discharge is difficult in comparision with stage measurement. In practice, there is often a need to infer the flow discharge from stage. In this example, the inflow hydrograph of a 2\,h flood in a synthetic compound channel, 1\,km long and 200\,m wide (500 cells), was inverted from the WSE recorded every 5\,min at three gauges along the main channel (Fig.~\ref{fig:hydrograph}a). The cross section of the compound channel has a flat main channel 40\,m wide, banks 2\,m high with a side slope of 1V:2.5H, and floodplains 75\,m wide on each side that rise at 3\,\% towards the valley walls. The streamwise bed slope is 0.001 and Manning's $n$ is 0.035\,s\,m$^{-1/3}$ everywhere. The downstream stage is held at 1.03\,m, approximately the normal depth of the base flow. The flood rises from 40 to 350\,m$^3$\,s$^{-1}$ and inundates the floodplains. The observations at the three gauges, i.e., the synthetic stage data, were generated with the explicit second-order model. In inversion, the unknown hydrograph was parameterised by its values at 15\,min intervals (9 values), and the inversion used the implicit backward Euler scheme with $\Delta t = 60$\,s and 25 iterations of L-BFGS. At the true knots themselves, the 60\,s step fails in one Newton solve during the first wetting of the bank; the inversion was not affected, as discussed in Sect.~\ref{sec:limitations}.

Starting from a constant base flow, the inversion recovered the hydrograph with an RMS error of 11.0\,m$^3$\,s$^{-1}$ and a peak of 356.7\,m$^3$\,s$^{-1}$, 1.9\,\% above the true peak (Fig.~\ref{fig:hydrograph}b, c). With 5\,mm of random noise added to the observations, the RMS error was 8.9\,m$^3$\,s$^{-1}$ and the peak 356.3\,m$^3$\,s$^{-1}$. The remaining errors are at the start of the rising limb, which the 15\,min resolution of the parameterisation cannot represent, and the loss was still decreasing after 25 iterations (Fig.~\ref{fig:hydrograph}d). Each inversion took about 10\,min on an NVIDIA A100 GPU. In earlier attempts with the explicit scheme, each loss evaluation needed about 27,000 time steps, and the dependence of the time step on the state made the loss slightly rough, so the line search struggled. The implicit backward Euler scheme with a fixed time step gave a much smoother loss with cheap gradients and was about ten times faster per iteration.

\subsection{Unsteady roughness calibration in the Savannah River}
\label{sec:savannah_unsteady}

\begin{figure*}[htp]
\includegraphics[width=17cm]{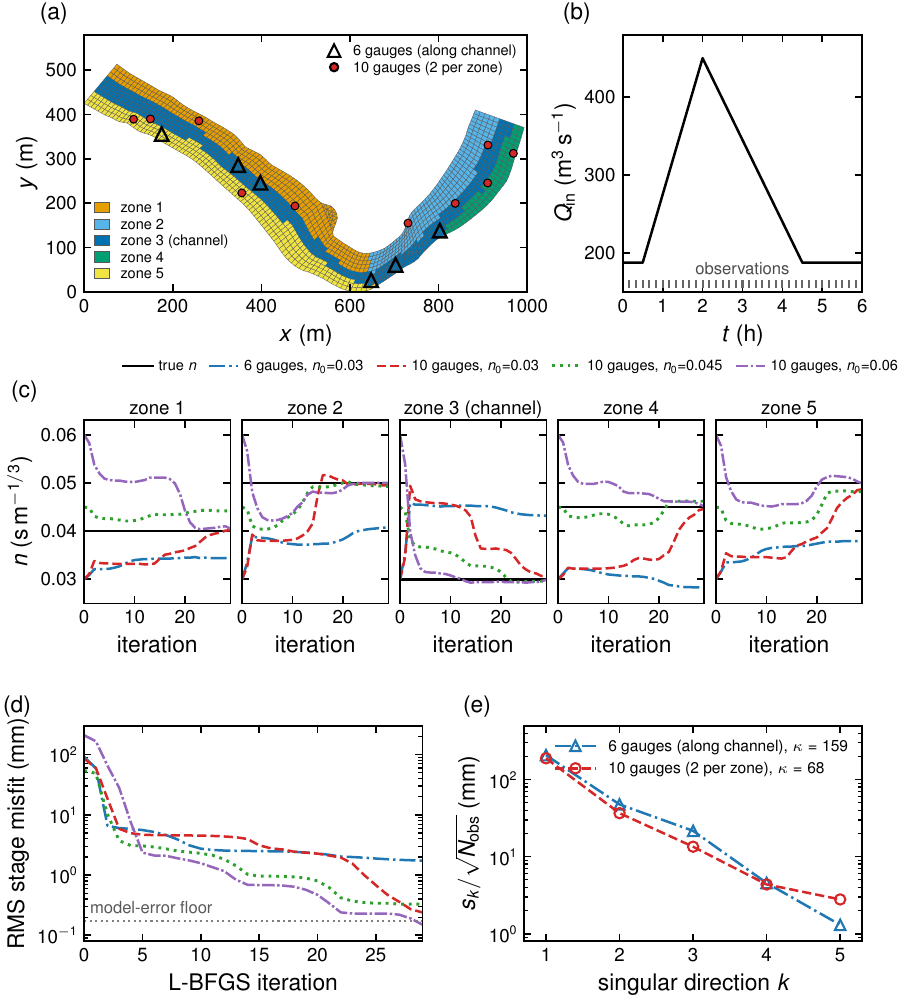}
\caption{Unsteady calibration of Manning's $n$ in the Savannah River: (a) roughness zones and two gauge layouts; (b) inflow hydrograph and observation times; (c) Manning's $n$ of each zone against L-BFGS iteration for six gauges and for ten gauges from three starting values $n_0$, with the true values; (d) RMS misfit of the stage against iteration, with the model-error floor; (e) singular values $s_k$ of the sensitivity matrix of the observations with respect to $\log n$, normalised by the square root of the number of observations $N_{\mathrm{obs}}$, with the condition numbers $\kappa$.}
\label{fig:savannah_unsteady}
\end{figure*}

Using the Savannah River case, this example combines the real unstructured mesh with an unsteady calibration and shows the role of the observation design. A flood rising from 187 to 450\,m$^3$\,s$^{-1}$ and back over 6\,h (Fig.~\ref{fig:savannah_unsteady}b) was simulated with the explicit second-order model and the true Manning's $n$, and the WSE was recorded every 10\,min. Two gauge layouts were tested: six gauges along the channel, and ten gauges with two in every roughness zone (Fig.~\ref{fig:savannah_unsteady}a). Manning's $n$ of the five zones was then inverted with the implicit backward Euler scheme of Sect.~\ref{sec:time} ($\Delta t = 30$\,s, 720 steps), a different integrator from the one that generated the data, and 30 iterations of L-BFGS. At the true $n$, the backward Euler scheme reproduces the observations to 0.17\,mm (RMS), which is the floor of the misfit.

Four inversions were run (Fig.~\ref{fig:savannah_unsteady}c, d): one with the six gauges, started from a uniform $n_0 = 0.03$\,s\,m$^{-1/3}$ in all zones, and three with the ten gauges, started from uniform $n_0 = 0.03$, 0.045 and 0.06\,s\,m$^{-1/3}$. The true values of the five zones lie between 0.03 and 0.05\,s\,m$^{-1/3}$, so the three starting values lie below, inside and above this range. All four inversions used the same observation times, integrator and number of L-BFGS iterations.

With six gauges, five of which lie in the main channel, the inversion fitted the observations to 1.74\,mm but missed $n$ by up to 44\,\% (Fig.~\ref{fig:savannah_unsteady}c, d). The data were fitted with compensating errors between zones. With ten gauges, the inversion recovered all five values to within 1.8\,\% and 0.3\,\% from two of three starting points (0.03 and 0.06\,s\,m$^{-1/3}$). The third start (0.045\,s\,m$^{-1/3}$) was still converging after 30 iterations, with an error of 10.5\,\%. 

The singular values of the sensitivity matrix explain the observed convergence behaviour (Fig.~\ref{fig:savannah_unsteady}e). The sensitivity matrix $\mathbf{S}$ has one row per observation (a gauge at one of the 36 observation times) and one column per roughness zone, and its entries are the derivatives of the observed stage with respect to $\log n$ of each zone; with 6 gauges, 36 observation times and 5 zones, $\mathbf{S}$ has $36 \times 6 = 216$ rows and 5 columns. Its singular value decomposition, $\mathbf{S} = \sum_k s_k \mathbf{u}_k \mathbf{v}_k^{\mathrm{T}}$, splits the five-dimensional parameter space into orthogonal directions $\mathbf{v}_k$, ordered by their singular values $s_1 \ge \dots \ge s_5$, with $\mathbf{u}_k$ the corresponding unit vectors in the space of observations. A change of $\log n$ by one unit along $\mathbf{v}_k$ changes the vector of observations by $s_k$ in the root-sum-square sense, i.e. by $s_k / \sqrt{N_{\mathrm{obs}}}$ per observation on average, where $N_{\mathrm{obs}}$ is the number of observations. A direction with a large singular value is well constrained by the data; a direction with a small singular value can be changed with little effect on the observations and is therefore poorly constrained. The ratio $\kappa = s_1 / s_5$, the condition number, measures how much better the best-constrained direction is resolved than the worst. $\mathbf{S}$ was evaluated at the true $n$ by one-sided finite differences of the backward Euler model ($\Delta t = 30$\,s), with steps of 0.02 in $\log n$, because the implicit solver provides reverse-mode derivatives only. The finite-difference matrix was checked against the adjoint: one reverse-mode gradient of $\mathbf{u}_k^{\mathrm{T}} \mathbf{y}$, where $\mathbf{y}$ is the vector of observed stages, gives $\mathbf{S}^{\mathrm{T}} \mathbf{u}_k$ exactly, and it agreed with $s_k \mathbf{v}_k$ to 0.8\,\% for the strongest direction of both gauge layouts and to 2\,\% (ten gauges) and 7\,\% (six gauges) for the weakest, within the truncation error of the one-sided differences.

The strongest direction $\mathbf{v}_1$ has components of the same sign in all zones: it changes the roughness of all zones together, i.e. the total resistance of the reach, and changes the stage by about 200\,mm per unit of $\log n$ (Fig.~\ref{fig:savannah_unsteady}e). The weakest direction $\mathbf{v}_5$ has components of mixed signs: it raises the roughness of some zones and lowers that of others, which redistributes the resistance between zones at nearly constant total resistance, and changes the stage by only 1.3\,mm with six gauges and 2.8\,mm with ten. The stage records at the gauges are therefore 70 to 160 times more sensitive to the total resistance of the reach than to its distribution among the zones.

This example illustrates two general points. First, exact gradients do not remove the non-convexity of calibration problems: several inversion starting points and multiple observations in every zone are needed. Second, sensitivities computed with the model identify such weak directions before the data are collected, which can guide the placement of gauges. Each inversion took 4 to 7\,h on an A100 GPU (six gauges) or 16 cores of a Xeon Gold 6342 CPU (ten gauges). Most of the time was spent in the line searches of L-BFGS.

\subsection{Evaluation of gradients}
\label{sec:gradient_checks}

All the inversion and closure learning examples used gradients computed by AD or the implicit function theorem. To verify that these gradients are correct, they were compared with central finite differences of the same discrete model on the cases of Sects.~\ref{sec:savannah_srh2d}, \ref{sec:verification}, \ref{sec:steady_ude} and \ref{sec:hydrograph} (Table~\ref{tab:gradients}). In every check, the differentiated quantity is a weighted sum, with fixed random weights, of the steady state or the final state over all cells, or, for the flood, of the depths saved every 15\,min, so that every knot of the hydrograph acts on it. All checks ran on 8 cores of an Intel Xeon Gold 6342 CPU in double precision.

The steady gradients from the implicit function theorem agree with finite differences to $3.4 \times 10^{-8}$ for the five Manning zones of the Savannah River and to $4.3 \times 10^{-8}$ for the bump. For the 22 weights of the closure network of Sect.~\ref{sec:steady_ude} at the start of training, the steady gradient agrees to better than $5 \times 10^{-8}$ for 20 of the 22 weights and to $9.9 \times 10^{-7}$ for the weight with the smallest gradient, and it costs about as much as the Newton solve itself (0.14\,s against 0.13\,s). Through the time-marching training configuration of the same case (adaptive Runge--Kutta integrator, 2,391 steps), forward-mode and reverse-mode sensitivities of the 22 weights agree with each other to $3.4 \times 10^{-5}$. Both agree with finite differences to between $10^{-5}$ and $3 \times 10^{-3}$, a limit set by the finite differences themselves, which sample runs at a tighter tolerance than the one the AD differentiates. Reverse mode cost 4.1 forward runs and forward mode 20.

For the flood of Sect.~\ref{sec:hydrograph}, the gradients with respect to the nine knots of the inflow hydrograph were checked through both unsteady integrators. Through 38,141 explicit steps, the agreement is between $2.7 \times 10^{-6}$ and $2.2 \times 10^{-3}$. The remaining difference comes from the CFL time-step selection, which depends on the state but is not differentiated. Through the implicit integrator with the settings of the inversion ($\Delta t = 60$\,s), the agreement is between $10^{-7}$ and $10^{-9}$ for the knots that act after the flood peak, but only $10^{-3}$ to 0.1 for the knots that act while the flood front advances. The latter is a limitation of the finite differences, not of the adjoint: with 60\,s steps, cells wet within a single step, so the loss is only piecewise smooth in those knots, and the finite-difference quotient does not converge as the perturbation is reduced. The implicit gradient cost 2.2 forward runs.

\begin{table*}[htp]
\caption{Gradient verification: relative difference between the gradient of each method and central finite differences of the same discrete model (second order). The cost is given in seconds for the steady solver and relative to one forward run of the same model otherwise.}
\label{tab:gradients}
\begin{tabular}{p{3.9cm}p{5.4cm}p{4.0cm}p{3.2cm}}
\hline
Method & Case & Relative difference & Cost of a gradient \\
\hline
Steady, implicit function theorem & Savannah River (Sect.~\ref{sec:savannah_srh2d}), 5 zonal Manning's $n$ & $\leq 3.4 \times 10^{-8}$ & 1.7\,s (Newton solve 1.4\,s) \\
Steady, implicit function theorem & Bump with hydraulic jump (Sect.~\ref{sec:verification}), Manning's $n$ & $4.3 \times 10^{-8}$ & 0.7\,s (Newton solve 0.7\,s) \\
Steady, implicit function theorem & Steady closure learning (Sect.~\ref{sec:steady_ude}), 22 NN weights & $\leq 9.9 \times 10^{-7}$ & 0.14\,s (Newton solve 0.13\,s) \\
Explicit SSP-RK3, 64 checkpoints & Flood of Sect.~\ref{sec:hydrograph}, 38,141 steps, 9 hydrograph knots & $2.7 \times 10^{-6}$ to $2.2 \times 10^{-3}$ & $\approx$\,7.3 forward runs \\
Implicit backward Euler, $\Delta t = 60$\,s & Same flood, inversion model, 9 hydrograph knots & $10^{-9}$ to $10^{-7}$; up to 0.11 during wetting (see text) & $\approx$\,2.2 forward runs \\
Forward mode against reverse mode & Transient closure learning (Sect.~\ref{sec:steady_ude}), 22 NN weights & $3.4 \times 10^{-5}$ & 20 and 4.1 forward runs \\
\hline
\end{tabular}
\end{table*}

\subsection{Computational performance}
\label{sec:performance}

\begin{figure}[htp]
\includegraphics[width=8.3cm]{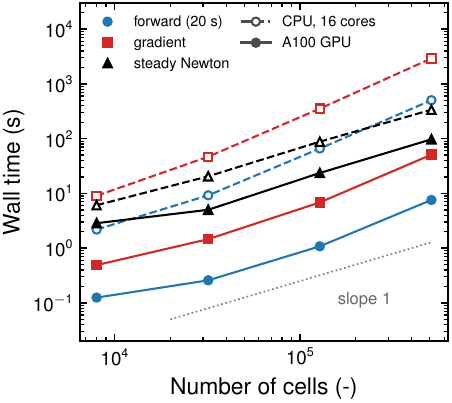}
\caption{Wall time of the forward run (20\,s simulated), the reverse-mode gradient (64 checkpoints) and the steady Newton solve against the number of cells, with the second-order scheme, on 16 CPU cores (open symbols, dashed lines) and an NVIDIA A100 GPU (filled symbols, solid lines).}
\label{fig:performance}
\end{figure}

The computational performance was measured on a straight channel, 1,000\,m long and 100\,m wide, with a bed slope of $10^{-4}$, a Gaussian bump 0.3\,m high at mid-length and Manning's $n = 0.03$\,s\,m$^{-1/3}$, meshed with 8,000 to 512,000 structured quadrilateral cells, with the second-order scheme in double precision. A constant discharge of 100\,m$^3$\,s$^{-1}$ enters at the inlet, the stage at the outlet is fixed at 2.0\,m, and the sides are walls. The forward run simulated 20\,s from rest with a horizontal free surface at 2.0\,m, with the adaptive Runge--Kutta integrator, and the gradient of the mean squared free-surface displacement at 20\,s with respect to Manning's $n$ was computed in reverse mode with 64 checkpoints. The steady Newton solver started from the same state and converged to the subcritical flow over the bump in six iterations at every mesh size. The CPU runs used 16 cores of an Intel Xeon Gold 6342 CPU, and the GPU runs one NVIDIA A100 GPU with 40\,GB of memory. All times exclude the compilation of the model (Table~\ref{tab:scaling} and Fig.~\ref{fig:performance}).

On the GPU, a forward run took 0.13\,s for 8,000 cells and 7.6\,s for 512,000 cells. Above about $10^5$ cells, the time grows in proportion to the number of cells multiplied by the number of time steps, which also increases as the cells become smaller. The GPU was 18 to 66 times faster than the 16 CPU cores for forward runs and 18 to 57 times faster for gradients, and its advantage grows with the mesh size. A reverse-mode gradient cost 3.9 to 6.7 forward runs on the GPU and 4.1 to 5.8 on the CPU.

The steady Newton solver took 6.1\,s for 8,000 cells and 337\,s for 512,000 cells on the CPU, and 2.9 and 97\,s when the residual and the Jacobian were evaluated on the GPU and the factorisation on its host CPU (12 cores of an Intel Xeon Gold 6248R). In both cases the sparse LU factorisation runs on the CPU and dominates the cost of large meshes. The gradient from the implicit function theorem added 9 to 26\,\% to the cost of the Newton solve. At 512,000 cells, it exceeded the memory of the GPU, which is a limitation of the present implementation, but it ran on the CPU in 424\,s including the Newton solve.

\begin{table*}[htp]
\caption{Wall time of the forward run (20\,s simulated), the reverse-mode gradient (64 checkpoints), the steady Newton solve and the Newton solve with the gradient from the implicit function theorem (IFT) for a channel with the second-order scheme. CPU: 16 cores of an Intel Xeon Gold 6342; GPU: NVIDIA A100 (40\,GB), with the sparse factorisations of the Newton solver on the host CPU of the GPU node (12 cores of an Intel Xeon Gold 6248R). Times are in seconds and exclude compilation. The IFT gradient for 512,000 cells exceeded the GPU memory.}
\label{tab:scaling}
\begin{tabular}{rrrrrrrrrr}
\hline
Cells & Steps & \multicolumn{2}{c}{Forward} & \multicolumn{2}{c}{Gradient} & \multicolumn{2}{c}{Newton} & \multicolumn{2}{c}{Newton + IFT} \\
 &  & CPU & GPU & CPU & GPU & CPU & GPU & CPU & GPU \\
\hline
8,000 & 104 & 2.19 & 0.125 & 9.05 & 0.490 & 6.08 & 2.86 & 6.96 & 3.31 \\
32,000 & 173 & 9.22 & 0.258 & 46.8 & 1.46 & 20.5 & 5.02 & 22.4 & 5.66 \\
128,000 & 296 & 65.9 & 1.08 & 353 & 6.84 & 88.0 & 23.6 & 102 & 27.9 \\
512,000 & 557 & 503 & 7.59 & 2,918 & 51.0 & 337 & 97.2 & 424 & -- \\
\hline
\end{tabular}
\end{table*}

\section{Discussion}
\label{sec:discussion}

\subsection{On the choice of a gradient method}

The applications show that no single gradient method is best for all problems, and that the choice matters more than the hardware. For steady flows, the implicit function theorem is the method of choice. A gradient costs about one Newton iteration and is independent of the number of parameters. It is also exact for the steady state itself. Differentiating through a time march to steady state gives the same gradient only when the march has converged to machine precision, at a cost that was two orders of magnitude higher in our tests (Sect.~\ref{sec:steady_ude}). In the author's opinion, many calibration studies that use steady observations, including the USWEs examples of v1.0 \citep{Liu2025}, would benefit from the approach of the implicit function theorem. 

For unsteady flows, the choice depends on the length of the event and the size of the time step. Explicit time stepping with checkpointing is robust at wet fronts, preserves positive depths and runs efficiently on GPUs. Its gradient costs a few forward runs, and its memory does not depend on the length of the simulation. The implicit scheme pays off when the time step can be much larger than the explicit stability limit. In our tests, one implicit step cost roughly as much as one hundred explicit steps, so the implicit scheme pays off only when its time step can be about a hundred times the explicit one. This is the case for long, slowly varying hydrographs and for meshes with very small cells, e.g. around structures. In these cases, its smoother loss and its cheap gradient (about 2 to 2.6 forward runs) made the hydrograph inversion of Sect.~\ref{sec:hydrograph} well behaved and about ten times faster per iteration than with the explicit scheme.

\subsection{On unstructured meshes}

Unstructured meshes do not prevent efficient execution on GPUs. Written as a gather of cell states, a computation per face and a sum per cell, the finite-volume method runs on GPUs with the same code as on CPUs. The exact Jacobian needs a number of colours that depends on the stencil but not on the mesh size (24 at second order for the Savannah River mesh). As a result, steady and implicit solves scale with the cost of the sparse factorisation rather than with the number of cells. The practical benefit is that \textit{Hydrograd} v2.0 works directly on the meshes and roughness zones that practitioners already build for many real-world applications. A calibrated model, or a learned closure, can therefore be transferred to production models without re-meshing.

\subsection{Scope and limitations}
\label{sec:limitations}

The model is fit for the calibration of flow resistance, the inversion of inflows, and the training of local closures in steady and unsteady 2D flows. Several limitations remain.

First, gradients across moving wet fronts and hydraulic jumps are those of the discrete model. They are exact for the discrete equations, but the discrete model is only piecewise smooth there. The hydrostatic reconstruction and the wet/dry treatment contain ``$\max$'' functions, and the explicit time step depends on the state. Losses that are dominated by the timing of wetting fronts may therefore be rough. The implicit scheme with a fixed time step is preferable for them.

Second, Newton's method may need many iterations, or fail at large time steps, where wet fronts advance quickly. The continuous blend to first order near dry cells reduced this problem substantially (Sect.~\ref{sec:discretisation}) but did not remove it. In the flood of Sect.~\ref{sec:hydrograph}, backward Euler diverged in one Newton solve when the flood first climbed the bank, at the true knots and at other, isolated hydrographs. In 13 hydrographs spaced evenly between the inverted and the true knots, 5 failed with $\Delta t = 60$\,s and 1 with $\Delta t = 30$\,s. The inversion, which started from a constant base flow, was not affected. All its loss values are finite and the gradient vanishes at the inverted knots. Their misfit is lower than that of the true knots at both step sizes ($6.6 \times 10^{-4}$ against $1.9 \times 10^{-3}$\,m$^2$ at 30\,s), because the 15\,min piecewise-linear hydrograph through the true samples is not the true hydrograph. In gradient mode, the implicit solver takes fixed steps without adaptively changing time step size, because the decision to adapt would make the loss discontinuous in the parameters. Steady states that do not exist, e.g. with an oscillating hydraulic jump or a moving wet front, cannot be found by the steady solver.

Third, the steady and implicit solvers factorise the Jacobian on the CPU. Their cost is dominated by the factorisation above about $10^5$ cells, and a GPU sparse direct solver or a preconditioned iterative solver would be needed for meshes of several million cells.

Fourth, calibration problems are often non-convex and poorly conditioned. As the Savannah River example showed (Sect.~\ref{sec:savannah_unsteady}), exact gradients do not guarantee that an optimiser finds the global minimum. Multiple starting points, observations in every roughness zone and complementary data types (e.g. velocities) remain important.

Fifth, the NN closures must be local functions of the flow variables, so that the Jacobian stays sparse and the learned closure is transferable. Closures that depend on non-local information, such as the distance traversed through a vegetation patch for the sheltering effect, must be supplied as additional local inputs.

\subsection{Future work}

The development of \textit{Hydrograd} will continue in four directions. First, GPU-based sparse linear solvers will remove the CPU bottleneck of the steady and implicit solvers. Second, sediment transport and morphodynamics will be added, so that bed evolution can be calibrated with the same gradient methods. Third, more closures will be added, such as a vegetation-drag closure, and the existing flow-resistance and turbulent-mixing closures will be trained on laboratory and field data. Finally, coupling with data assimilation will allow the model to be used in operational settings.

\conclusions
\label{sec:conclusions}

\textit{Hydrograd} v2.0 is a differentiable 2D shallow-water model that works on the unstructured meshes of production models, handles wetting and drying, and provides gradients whose cost matches the type of problem. Steady problems are solved by Newton's method with an exact sparse Jacobian, and their gradients follow from the implicit function theorem at the cost of one linear solve. Unsteady problems are differentiated in reverse mode with checkpointing or with an implicit time integrator. The solution and its gradients were verified against analytical solutions and finite differences with high accuracy.

The applications lead to three conclusions. First, for steady observations the steady solver with the implicit function theorem should replace differentiation through time marching. It was up to two orders of magnitude faster in closure learning, and its gradient should only depend on the steady state itself rather than of an incompletely converged transient. Second, for unsteady problems the explicit and implicit integrators are complementary. The explicit scheme is robust at wet fronts and efficient on GPUs, and the implicit scheme gives a smoother loss and a gradient of about two forward runs when its time step can be much larger than the explicit limit, as in the inversion of an inflow hydrograph. Third, exact gradients do not remove the ill-conditioning of calibration problems. Stage data constrained the total resistance of a river reach far more strongly than its distribution among zones. The design of the observations, including additional data types such as velocities, are as important as the optimiser.

\codedataavailability{The source code of \textit{Hydrograd} v2.0, the user manual, the input files of all examples, and the scripts that reproduce every figure and table of this paper are archived on Zenodo at \url{https://doi.org/10.5281/zenodo.24045388} (version 2.0.0, the exact version described in this paper; all versions: \url{https://doi.org/10.5281/zenodo.24045387}) and are distributed under the MIT licence. The development version is available at \url{https://github.com/psu-efd/Hydrograd}. \textit{Hydrograd} v1.0 (Hydrograd.jl) is available at \url{https://github.com/psu-efd/Hydrograd.jl}. The model requires Python 3.11 or newer with JAX; it has been tested on Linux on Intel Xeon CPUs and NVIDIA A100 GPUs.}

\appendix

\section{Differences from v1.0}
\label{app:differences}

Table~\ref{tab:differences} lists the differences between \textit{Hydrograd} v2.0 and v1.0 (Hydrograd.jl) \citep{Liu2025}. 

\begin{table*}[htp]
\caption{Differences between \textit{Hydrograd} v1.0 (Hydrograd.jl) and v2.0.}
\label{tab:differences}
\begin{tabular}{p{4.6cm}p{5.4cm}p{5.4cm}}
\hline
Item & \textit{Hydrograd} v1.0 (Hydrograd.jl) & \textit{Hydrograd} v2.0 \\
\hline
Default scheme & Roe scheme on the free-surface elevation, wet domains & Hydrostatic reconstruction and HLLC, wet/dry, first or second order; the Roe scheme remains an option \\
Distribution of the inflow over inlet faces & conveyance $L_f^{5/3} h / n$ & conveyance $L_f h^{5/3} / n$ (corrected) \\
Still-water depth when the bed is inverted & $h_s$ from the initial bed & $h_s = w_s - z_b(\boldsymbol{\theta})$, consistent with the trial bed during bathymetry inversion \\
Normalisation layer of the NN closure & over all cells (non-local) & none by default; per cell as an option (local closure) \\
Velocity in the loss & $q/(h + \sqrt{\epsilon_{\mathrm{mach}}})$ & $q/(h + h_{\mathrm{small}})$, consistent with the truth files \\
$n(h, |\mathbf{u}|, k_s)$ law at very small Reynolds number & undefined (not a number) & bounded at the limits of validity \\
Steady states & time marching & Newton iteration with gradients from the implicit function theorem \\
Long unsteady runs & reverse mode through all steps & checkpointing, or implicit time stepping with per-step adjoints \\
\hline
\end{tabular}
\end{table*}

\noappendix

\authorcontribution{XL conceived the model, developed the software, carried out the simulations and analyses, and wrote the paper.}

\competinginterests{The author declares that there is no conflict of interest.}

\begin{acknowledgements}
Computations for this research were performed on the Pennsylvania State University's Institute for Computational and Data Sciences' Roar supercomputer. The author used Claude Code (Anthropic) as an assistant in software development (porting from Julia, testing and refactoring the code) and proofreading the text. The author conceived the research idea, designed the work, checked all code, results, and wrote the manuscript draft.
\end{acknowledgements}

\bibliographystyle{copernicus}
\bibliography{references}

\end{document}